\documentclass[10pt,aps,prb,twocolumn,superscriptaddress,amssymb,floatfix]{revtex4}
\usepackage{graphicx}
\usepackage{epstopdf}
\usepackage[T1]{fontenc}
\usepackage[latin9]{inputenc}
\usepackage{amsbsy}
\begin{document}


\title{Probing the superconducting ground state of the noncentrosymmetric superconductors Ca\textit{T}Si$_3$ (\textit{T} = Ir, Pt) using muon-spin relaxation and rotation}


\author{R. P. Singh}
\affiliation{Department of Physics, University of Warwick, Coventry, CV7 7AL, UK}

\author{A. D. Hillier}
\affiliation{ISIS facility, STFC Rutherford Appleton Laboratory, Harwell Science and Innovation Campus, Oxfordshire, OX11 0QX, UK}
\author{D. Chowdhury}
\affiliation{ISIS facility, STFC Rutherford Appleton Laboratory, Harwell Science and Innovation Campus, Oxfordshire, OX11 0QX, UK}

\author{J. A. T. Barker}
\affiliation{Department of Physics, University of Warwick, Coventry, CV7 7AL, UK}
\author{D. M$^\mathrm{c}$K. Paul}
\affiliation{Department of Physics, University of Warwick, Coventry, CV7 7AL, UK}
\author{M. R. Lees}
\affiliation{Department of Physics, University of Warwick, Coventry, CV7 7AL, UK}
\author{G. Balakrishnan}
\affiliation{Department of Physics, University of Warwick, Coventry, CV7 7AL, UK}

\date{\today}

\begin{abstract}
The superconducting properties of Ca$T$Si$_3$ (where \textit{T} = Pt and Ir) have been investigated using muon spectroscopy. Our muon-spin relaxation results suggest that in both these superconductors time-reversal symmetry is preserved, while muon-spin rotation data show that the temperature dependence of the superfluid density is consistent with an isotropic $s$-wave gap. The magnetic penetration depths determined from our transverse-field muon-spin rotation spectra are found to be 448(6) and 150(7)~nm for CaPtSi$_3$ and CaIrSi$_3$ respectively.
\end{abstract}

\pacs{74.20.Rp, 74.70.Dd}
\keywords{}

\maketitle
\section{Introduction}
There has been a great deal of interest in noncentrosymmetric superconductors (NCS), instigated by the discovery of the noncentrosymmetric heavy-fermion superconductor CePt$_3$Si~\cite{Bauer04}. The absence of a center of inversion in the crystal structure along with a nontrivial anti-symmetric spin-orbit coupling, leads to the intriguing possibility of a superconducting state with an admixture of spin-triplet and spin-singlet pairs. Despite intense theoretical and experimental efforts~\cite{Gorkov01,Frigeri04,Frigeri06,Sigrist08,Bauer12} the study of the physics of noncentrosymmetric superconductors remains a dynamic and active field.

One manifestation of unconventional superconductivity is the breaking of time-reversal symmetry (TRS). The magnetic moments associated with the Cooper pairs are non-zero for such superconductors. A local alignment of these moments produces spontaneous, but extremely small, internal magnetic fields. Muon-spin relaxation is especially sensitive to small changes in internal fields and can easily measure fields of $\approx0.1$~G which correspond to moments that are just a few hundredths of a $\mu_\mathrm{B}$. Muon-spin relaxation is thus an ideal probe with which to search for TRS breaking~\cite{Schenck,Lee98,Sonier07,Yaouanc}. Time-reversal symmetry breaking is rare and is only observed directly in a few superconductors, e.g. Sr$_2$RuO$_4$,\cite{Luke98}  UPt$_3$,\cite{Luke93} (although not without controversy),\cite{deReotier95} (U,Th)Be$_{13}$,\cite{Heffner90} PrOs$_4$Sb$_{12}$,\cite{Aoki03} LaNiC$_2$,\cite{Hillier09} PrPt$_4$Ge$_{12}$,\cite{Maisuradze10} (PrLa)(OsRu)$_4$Sb$_{12}$,\cite{Shu11} LaNiGa$_2$,\cite{Hillier12} and more recently SrPtAs~\cite{Biswas13}. The possibility of mixed spin-singlet spin-triplet pairing in noncentrosymmetric superconductors makes them prime candidates to exhibit TRS breaking~\cite{Bauer12}. To date, the only truly NCS reported to show TRS breaking are LaNiC$_2$ and Re$_6$Zr,\cite{Hillier09,Singh14} however, for LaNiC$_2$, a mixing of singlet and triplet states is forbidden due to the symmetry of the structure~\cite{Quintanilla10}, but is allowed for Re$_6$Zr.

Many other NCS have been studied by magnetization, transport, and heat capacity measurements  e.g. Nb$_{0.18}$Re$_{0.82}$,\cite{Karki201a} LaRhSi$_3$,\cite{Anand11} Mg$_{10}$Ir$_{19}$B$_{16}$,\cite{Klimczuk2007a} Mo$_{3}$Al$_{2}$C\cite{Karki2011a} and Re$_{3}$W,\cite{Biswas11} and several have been shown to exhibit unconventional superconducting behavior including features such as upper critical fields close to the Pauli limit, evidence for multiple gaps, or a significant admixture of a triplet component to the superconducting order parameter. Muon spectroscopy studies have been performed on some of these compounds e.g. Re$_3$W,  LaRhSi$_3$, LaPtSi$_3$, LaPdSi$_3$~\cite{Biswas11,Anand11,Smidman14}. However, no spontaneous fields were observed in the superconducting state in any of these materials. This indicates that TRS breaking is either undetectable or not present in the superconducting state of these compounds.

Muon-spin rotation can be used to accurately determine the magnetic penetration depth $\lambda$ and hence the temperature dependence of the superfluid density, yielding information on the symmetry of the superconducting gap~\cite{Schenck,Lee98,Sonier07,Yaouanc}. This technique has been successfully applied to the study of a number of NCS including Re$_{3}$W\cite{Biswas11} and LaPtSi$_3$~\cite{Smidman14}. Using muon-spin relaxation and rotation ($\mu$SR) together can provide crucial information on the nature and mechanisms of superconductivity in new materials.

\begin{figure}
\includegraphics[width=6.cm,angle=270,origin=c]{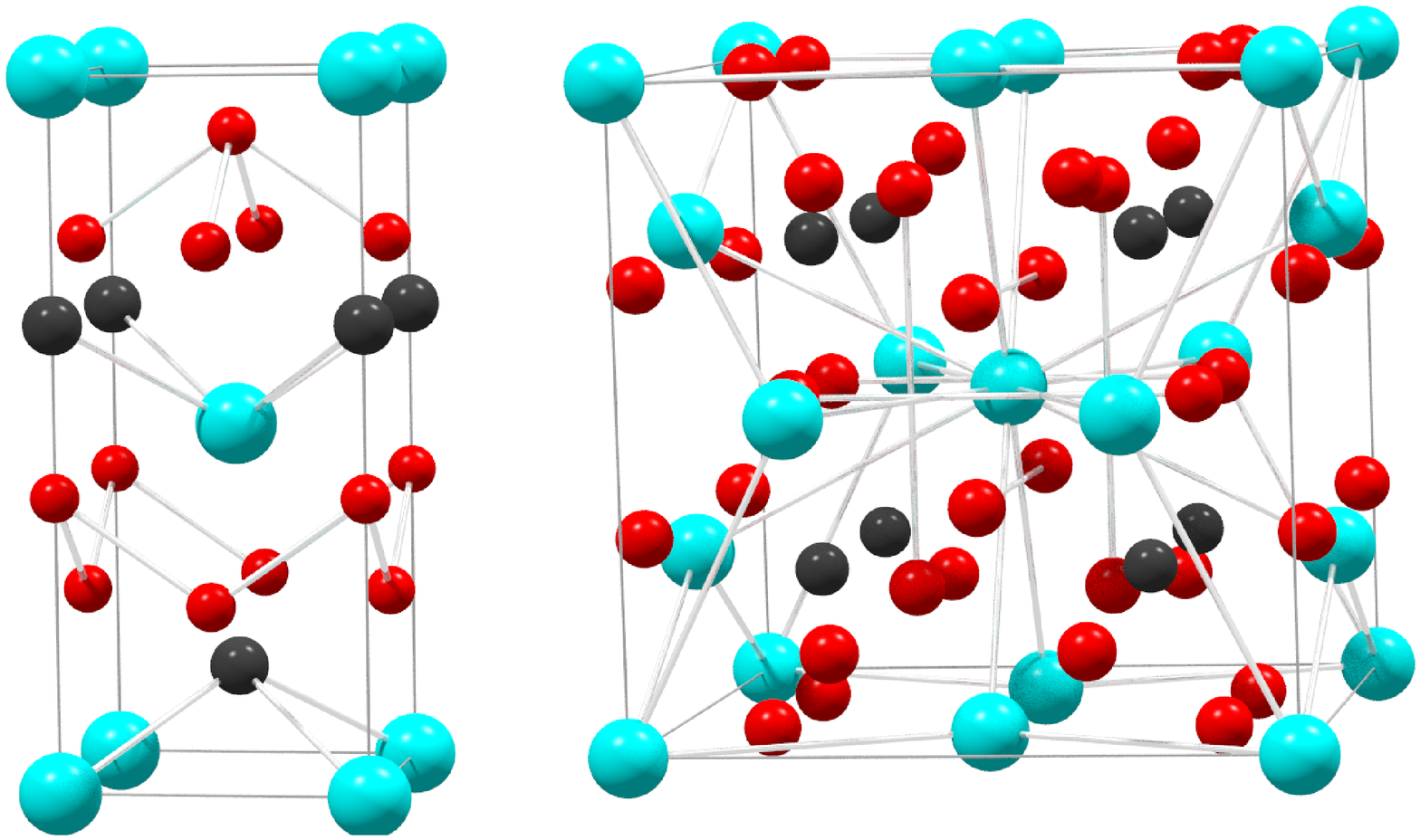}
\caption{\label{fig1:struct} (color online) Two possible crystal structures of the $RTX_3$ (113) compounds. Left is the noncentrosymmetric structure with the space group $I4mm$ and right is the centrosymmetric structure with the space group $Pm\bar{3}n$. The light (blue) spheres are the $R$ atoms, the black spheres the $T$ atoms, and the dark (red) spheres the $X$ atoms. }
\end{figure}

Recently a new  family of superconducting materials has been discovered with the formula \textit{RTX}$_3$ (where \textit{R}~=~rare earth, \textit{T}~=~transition metal, and \textit{X}~=~Si or Ge). These 113 materials can crystallize in either the noncentrosymmetric BaNiSn$_3$-type crystal structure (space group $I4mm$), or in the centrosymmetric LaRuSn$_3$-type cubic structure (space group $Pm\bar{3}n$) shown in Fig.~\ref{fig1:struct}. Several 113 materials with a noncentrosymmetric structure such as CeCoSi$_3$~\cite{Haen85,Iwamoto95}, CeRhSi$_3$~\cite{Kimura05,Kimura07}, and CeIrSi$_3$~\cite{Sugitani06,Mukuda08} exhibit novel ground states. For example, at ambient pressure CeCoSi$_3$ becomes superconducting at 1.3~K, while CeRhSi$_3$ and CeIrSi$_3$ exhibit antiferromagnetic ordering at 1.5 and 5~K, respectively. The latter two compounds also reveal superconductivity under applied pressure. In CeRhSi$_3$ the superconducting transition temperature $T_c$ increases from 0.45  to 1.1~K between 0.4 and 2.3~GPa\cite{Kimura05,Kimura07}, while for CeIrSi$_3$ $T_c$ increases from 0.5  to 1.6~K between 1.8 and 2.5~GPa~\cite{Sugitani06}.  In these compounds, the competition between the on-site (nonmagnetic) Kondo interaction and the oscillatory inter-site long-range magnetic interactions, known as Rudermann-Kittel-Kasuya-Yosida (RKKY) interaction, plays an important role in the observed and rather unusual physical properties. Replacing Ce with a non-magnetic analogue such as La also produces superconductors which exhibit unusual properties. For example, LaRhSi$_3$ and LaPdSi$_3$ are type I superconductors, whereas LaPtSi$_3$ is type II~\cite{Anand11,Smidman14}.

In this paper we report on the properties of the new $f$-electron free 113 compounds, CaIrSi$_3$ and CaPtSi$_3$, both of which have the potential to be of great interest. CaIrSi$_3$ and CaPtSi$_3$ are NCS with superconducting transition temperatures of 3.6 and 2.3~K respectively and therefore do not require pressure to induce the superconducting state unlike their Ce analogues~\cite{Eguchi11,Eguchi_sces}. Specific heat data are in general agreement with these superconductors being fully gapped, however, CaIrSi$_3$ appears to show a deviation from a pure $s$-wave gap, which has been suggested as evidence for a multiband or anisotropic gap~\cite{Eguchi11}. In the presence of Ir and Pt it is expected that spin-orbit coupling will be significant, strengthening the possibility that the mechanisms for superconductivity might not be entirely conventional in these materials. Therefore, it is timely and interesting to probe the superconducting state of CaIrSi$_3$ and CaPtSi$_3$ using muon spectroscopy. In this work, muon-spin relaxation is used to search for evidence of TRS breaking in these two superconductors. Muon-spin rotation is used to determine the temperature dependence magnetic penetration depth. Since $\lambda(T)$ is directly related to the superfluid density, the pairing symmetry can then be determined. 

\section{Experimental Details}
\subsection{Sample preparation}
Polycrystalline samples of CaPtSi$_3$ and CaIrSi$_3$ were prepared by arc melting stoichiometric quantities of high purity Ca ($5\%$ excess of Ca to compensate for any weight loss), Pt/Ir and Si in a tri-arc furnace under an argon (5N) atmosphere on a water-cooled copper hearth. In order to minimize the loss of the Ca by evaporation, melting is done in two steps. In the first step, Pt/Ir are melted with Si. The observed weight loss during the melting of binaries Pt/Ir-Si is negligible. In the second step, Pt/Ir-Si binaries are melted with $5\%$ excess of Ca. The sample buttons were melted and flipped several times to improve phase homogeneity. 

\subsection{Sample characterization}
Powder x-ray diffraction data were collected for both samples. Refinement of the x-ray data (see table~\ref{table_of_structure}) confirmed both the samples had the tetragonal structure (space group $I4mm$ (No. 107)) with lattice parameters which are in good agreement with those reported earlier~\cite{Eguchi11}. There are some impurity peaks present in both the CaIrSi$_3$ and CaPtSi$_3$ samples at the same positions observed by Eguchi~\textit{et al}~\cite{Eguchi11}.

\begin{table}
\caption{Lattice parameters of noncentrosymmetric CaPtSi$_3$ and CaIrSi$_3$ determined from powder x-ray diffraction data collected at 298~K.}
\label{table_of_structure}
\begin{center}
\begin{tabular}[b]{lll}\hline\hline
{}~~~~~~~~&CaPtSi$_3$~~~~~~~~&CaIrSi$_3$\\\hline
Structure & Tetragonal & Tetragonal\\
Space group~~~~~~~~& $I4mm$ & $I4mm$\\
\textit{a} (nm) & 0.42182(5) & 0.41957(2)\\
\textit{c} (nm) & 0.9880(2) & 0.98711(7)\\\hline\hline
\end{tabular}
\par\medskip\footnotesize
\end{center}
\end{table} 

\begin{figure}
\includegraphics[width=7.cm]{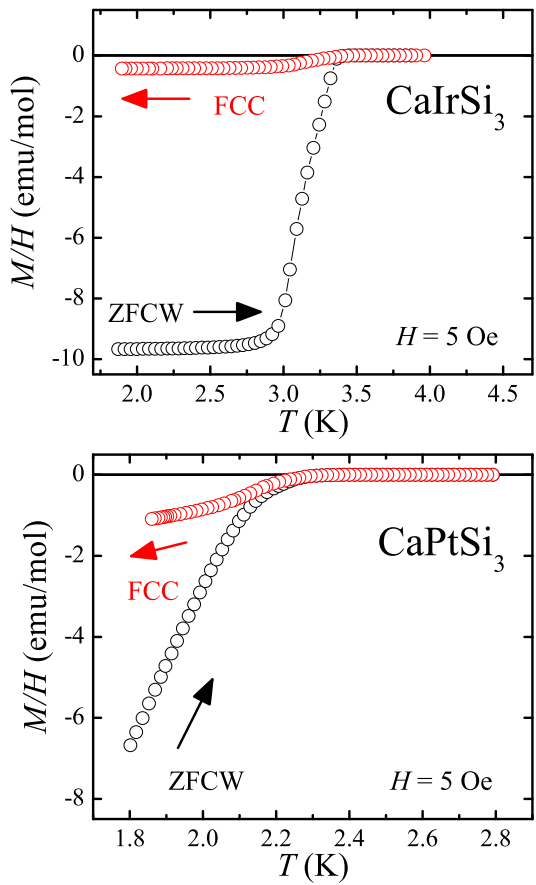}
\caption{\label{fig2:mag}(color online) Temperature dependence of the dc magnetization for CaIrSi$_3$ (upper) and CaPtSi$_3$ (lower). The samples were cooled in zero-field and a field of 5~Oe was then applied. Data were collected on zero-field-cooled warming (ZFCW) and during a subsequent field-cooled cooling (FCC) in the same applied field.}
\end{figure}

In order to confirm the superconducting transition temperatures of the samples, dc magnetic susceptibility measurements were made using a Quantum Design Magnetic Property Measurement System. Fig.~\ref{fig2:mag} shows the magnetic susceptibility as a function of temperature in an applied field of 5~Oe. The observed superconducting transition temperatures $T_{C}$ for CaPtSi$_3$ and CaIrSi$_3$ are approximately 2.3 and 3.5~K respectively (see Table~\ref{table_of_SC_parameters}). These transition temperatures are in good agreement with previously reported results measured by dc susceptibility on samples with the same composition~\cite{Eguchi11}. There is no evidence from the dc susceptibility data that the impurities present in our samples order magnetically or become superconducting. Since muon spectroscopy probes the full volume of the sample the results presented below are representative of the majority superconducting phases. 

\subsection{Muon spectroscopy}
The muon-spin relaxation measurements in zero field (ZF) and muon-spin rotation experiments in transverse fields (TF), were carried out at the ISIS pulsed neutron and muon facility using the MuSR spectrometer~\cite{King14}. At the ISIS facility, a pulse of protons with a full width at half maximum of $\approx 70$~ns are produced every 20~ms, with 4 out of 5 pulses going through the muon target. The muons produced are implanted into the sample and decay with an average lifetime of $2.2~\mu$s into a positron which is emitted preferentially in the direction of the muon spin axis along with two neutrinos. These positrons are detected and time stamped in the 64 detectors which are positioned either before, $F$, or after, $B$ the sample for longitudinal (relaxation) experiments. The asymmetry $A$ of the $\mu$SR time spectrum is then obtained as $A(t)=(F(t)-\alpha B(t))/(F(t)+\alpha B(t))$, where $\alpha$ represents a relative counting efficiency of the forward and backward detectors~\cite{Lee98,Yaouanc}. Using these counts the asymmetry in the positron emission can be determined and, therefore, the muon polarization is measured as a function of time. 

For the transverse-field experiments, a magnetic field is applied perpendicular to the initial muon spin direction and momentum. In this configuration, the signals from the instruments 64 detectors are normalized and reduced to two orthogonal components which are then fitted simultaneously.

Powder samples of each material were mixed with GE varnish and mounted onto silver holders. Any muons which stop in silver give a time independent background for ZF-$\mu$SR experiments and a non-decaying precession signal in the TF-$\mu$SR. The sample holder and sample were mounted into a helium-3 cryostat with a temperature range of 0.3 to 50~K. The samples were cooled to base temperature in zero field and the relaxation spectra were collected at fixed temperature upon warming while still in zero field. The stray fields at the sample position were canceled to within 10~mG by a flux-gate magnetometer and an active compensation system controlling three pairs of correction coils. The TF-$\mu$SR experiments were conducted in a range of applied fields from 50 to 600~Oe. The field was applied above the superconducting transition before cooling.

\section{Results and Discussion}
\subsection{Zero-field muon-spin relaxation}
\begin{figure}
\includegraphics[width=7.cm]{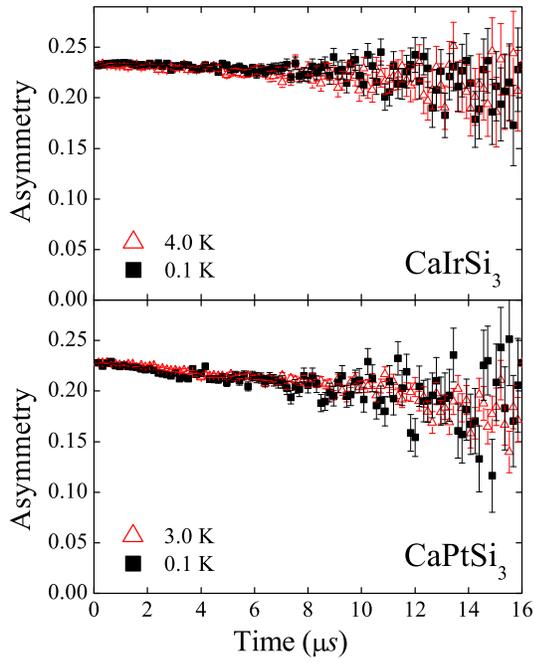}
\caption{\label{fig3:ZFspectra} (color online) Zero-field muon-spin relaxation spectra for CaIrSi$_3$ (upper panel) and  CaPtSi$_3$ (lower panel) at temperatures above (open symbols) and below (closed symbols) $T_c$.}
\end{figure}

Firstly, let us consider the zero-field muon-spin relaxation results (see Fig.~\ref{fig3:ZFspectra}). The absence of an oscillation in the ZF-$\mu$SR asymmetry data at all temperatures for both samples confirms that there are no coherent magnetic fields, usually, associated with long-range magnetic order. In the absence of atomic moments, in Ca\textit{T}Si$_3$ the muon-spin relaxation is expected to arise from the local fields associated with nuclear moments. These moments are usually static on the timescale of the muon and are randomly orientated. In a case such as this the depolarization function can be described by a Kubo-Toyabe function~\cite{Lee98,Yaouanc}. In Fig.~\ref{fig3:ZFspectra}, we can see that for both CaIrSi$_3$ and CaPtSi$_3$ the data is relatively flat and does not have the characteristic shape of the aforementioned Kubo-Toyabe function. This indicates that the fields from the nuclear moments are small. Moreover, the $\mu$SR signals for temperatures above and below the superconducting transition overlay and the depolarization rate is the same. This indicates that time-reversal symmetry is preserved, as would be expected in a conventional singlet superconductor, or at least any symmetry breaking field is not observable by $\mu$SR.

\begin{figure}
\includegraphics[width=7.cm]{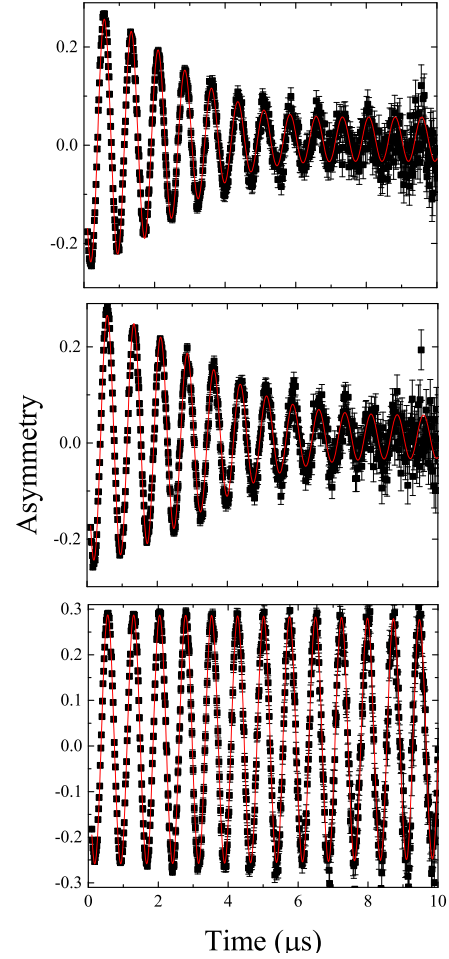}
\caption{\label{fig4:TFspectra} Typical muon-spin rotation spectra for CaPtSi$_3$ in a transverse field of 100~Oe at temperatures of 0.2~K (upper), 1.1~K(middle) and 1.95~K (lower). The lines are fits to the data using Eq.~\ref{eqn1:fitfunctf} as described in the text.}
\end{figure}

\subsection{Transverse-field muon-spin rotation}
Transverse-field muon-spin rotation can be used to determine the magnetic penetration depth $\lambda$. Figure~\ref{fig4:TFspectra} shows typical spectra with a transverse applied field of 100~Oe at $T=0.2$, 1.1, and 1.95~K after being cooled through $T_c$. The TF-$\mu$SR spectra were fit as a sum of sinusoidally oscillating components, each within a Gaussian relaxation envelope,
\begin{equation}
G_x(t)=\sum_{i=1}^{n} A_i\exp\left(-\frac{\sigma_i^2t^2}{2}\right)\cos\left(\gamma_{\mu}B_i t+\varphi\right),
\label{eqn1:fitfunctf}
\end{equation}

\begin{figure}
\includegraphics[width=7.cm]{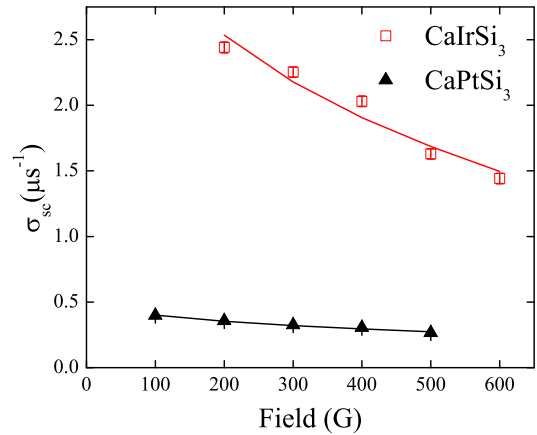}
\caption{\label{fig5:svb} Field dependence of the muon depolarization rate at $T=100$~mK for CaIrSi$_3$ (open symbols) and CaPtSi$_3$ (closed symbols). The lines are fits to the data using Eq.~\ref{eqn2:Brandt} as described in the text.}
\end{figure}

\noindent where $A_i$ is the initial asymmetry, $\sigma_i$ is the Gaussian relaxation rate, and $B_i$ is the first moment for the $i$-th component of the field distribution~\cite{Maisuradze09,Weber}. There is a common phase offset $\varphi$ and $\gamma_{\mu}$ is the muon gyromagnetic ratio. In these fits, $\sigma_n$ for the $n$-th component is set to zero and corresponds to a background term arising from those muons which are implanted into the silver sample holder producing an oscillating signal that has no depolarization, as silver has a negligible nuclear moment. Using Eq.~\ref{eqn1:fitfunctf} is equivalent to assuming a field distribution $P\left(B\right)$ within the sample given by

\begin{equation}
P(B)=\gamma_{\mu}\sum_{i=1}^{n-1} \frac{A_i}{\sigma_i}\exp\left(-\frac{\gamma^{2}_{\mu}\left(B-B_i\right)^2}{2\sigma^2_i}\right).
\label{eqn1:PB}
\end{equation}

\noindent The second moment of the field distribution within the sample $\left\langle \Delta B^2\right\rangle$ is

\begin{equation}
\left\langle \Delta B^2\right\rangle=\left(\frac{\sigma}{\gamma_{\mu}}\right)^2=\sum_{i=1}^{n-1} \frac{A_i}{A_{\mathrm{tot}}}\left[\left(\frac{\sigma_i}{\gamma_{\mu}}\right)^2+\left(B_i-\left\langle B\right\rangle\right)^2\right],
\label{eqn1:secondmoment}
\end{equation}

\noindent where $A_{\mathrm{tot}}=\sum_{i=1}^{n-1} A_i$ and $\left\langle B\right\rangle=\sum_{i=1}^{n-1} \frac{A_iB_i}{A_{\mathrm{tot}}}$. The superconducting component of the second moment $\sigma_\mathrm{sc}$ is then given by $\sigma^2_\mathrm{sc}=\sigma^2-\sigma^2_\mathrm{nm}$ where $\sigma^2_\mathrm{nm}$ is the signal in the normal state due to the nuclear moments.

The spectra from the CaIrSi$_3$ were best described by three oscillating functions whereas the spectra from the CaPtSi$_3$ could be described by just two~\cite{Maisuradze09}. The field dependence of the superconducting depolarization rates $\sigma_\mathrm{sc}$ are shown in Fig.~\ref{fig5:svb}. As the field increases the depolarization rates decrease as may be expected for a superconductor when the applied field is a significant fraction of the upper-critical field $B_{\mathrm{c2}}$~\cite{Sonier07}. The field dependence of $\sigma_\mathrm{sc}$ can be used to determine the magnetic penetration depth and to give an estimate for the upper-critical field (see Fig.~\ref{fig5:svb}). $B_{\mathrm{c2}}$ can be independently verified using other measurements. The $\sigma_\mathrm{sc}\left(B\right)$ data shown in Fig.~\ref{fig5:svb} were fit using Eq.~\ref{eqn2:Brandt} 


\begin{equation}
\sigma_{sc}\left[\mu s^{-1}\right]=A\times (1-b)(1+1.21(1-\sqrt{b})^3)\lambda^{-2}\left[\mathrm{nm}\right],
\label{eqn2:Brandt}
\end{equation} 

\noindent where $\lambda$ is in nm, $b = B/B_{\mathrm{C2}}$ is the ratio of applied field to the upper critical field, and $A$ is a prefactor related to the structure of the flux-line lattice ($A=4.83\times10^4$ for a hexagonal lattice)~\cite{Brandt88,Brandt03}. Assuming the penetration depth follows either a two-fluid model $\left(\lambda^{-2}\left(T\right)/\lambda^{-2}\left(0\right)=\left[1-\left(T/T_{c}\right)^4\right]\right)$ or can be described using the local (London) approximation for an $s$-wave gap superconductor (see below) gives penetration depths $\lambda(0)$ of 448(6) and 150(7)~nm for CaPtSi$_3$ and CaIrSi$_3$ respectively. The upper critical fields estimated from the $\sigma_\mathrm{sc}$ versus $B$ data and from extrapolations to zero kelvin of the $B_{\mathrm{c2}}\left(T\right)$ curves determined from $M\left(H\right)$ loops collected at temperatures above 1.5~K (data not shown) are in good agreement with those reported by Eguchi~\textit{et al}.~\cite{Eguchi11} from magnetic and transport data, although as in Ref.~\onlinecite{Eguchi11} there is considerable uncertainty associated with these estimates. More comprehensive data sets down to low temperatures are required to accurately determine $B_{\mathrm{c2}}(0)$ for both materials.

\begin{figure}
\includegraphics[width=7.cm]{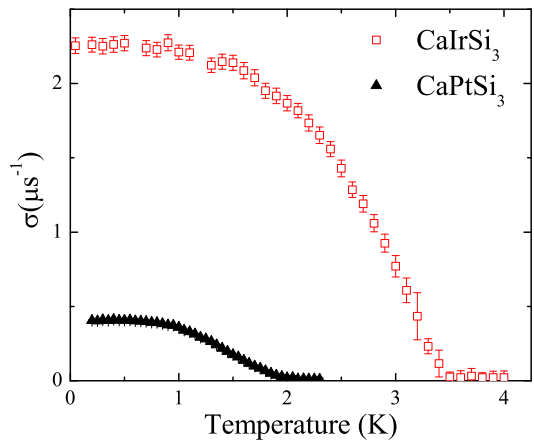}
\caption{\label{fig6:svt} Temperature dependence of the muon depolarization rate $\sigma$ for CaPtSi$_3$ (closed symbols) and CaIrSi$_3$ (open symbols) .}
\end{figure}

\begin{figure}
\includegraphics[width=7cm]{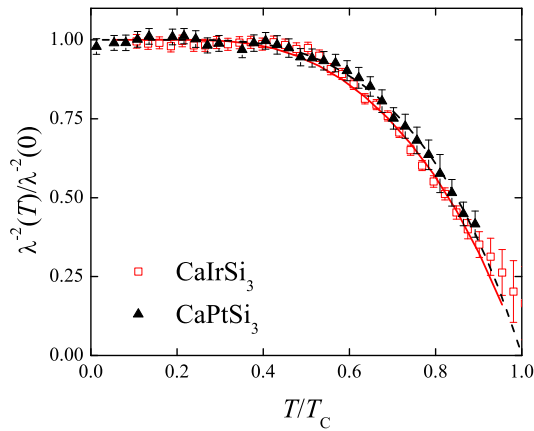}
\caption{\label{fig7:lvt} Normalized inverse square of the London penetration depth (the superfluid density) $\lambda^{-2}\left(T\right)/\lambda^{-2}\left(0\right)$ versus the reduced temperature $T/T{\mathrm{c}}$ for CaPtSi$_3$ and CaIrSi$_3$. The lines are fits to the data as described in the text. The closed symbols and solid line are the data and fit for the CaPtSi$_3$, the open symbols and dashed line are the data and fit for CaIrSi$_3$.}
\end{figure}

The temperature dependence of the muon depolarization rate $\sigma$ for CaPtSi$_3$ and CaIrSi$_3$ are given in Fig.~\ref{fig6:svt}. For both samples, the data shows a plateau and then decrease as the temperature is increased. $\sigma$ then levels off at a temperature slightly less than $T_c$. The depolarization rates at higher temperatures ($T\geq T_c$) are small, in agreement with the zero-field data discussed above. This depolarization $\sigma_{\mathrm{nm}}$ is associated with the nuclear moments. Below $T_c$, the depolarization rates are related to the magnetic penetration depth (see Eq.~\ref{eqn2:Brandt}) and therefore the structure of the superconducting gaps for the two materials can be investigated. After subtracting $\sigma_{\mathrm{nm}}$ from $\sigma$ to give $\sigma_{\mathrm{sc}}$ as described above, $\lambda$ can be calculated at each temperature, with a correction for the strong field dependence of the depolarization rates made using the $B_{\mathrm{c2}}(T)$ data from Ref.~[\onlinecite{Eguchi11}] and Eq.~\ref{eqn2:Brandt}. The temperature dependence of $\lambda\left(T\right)$ determined in this way for CaPtSi$_3$ and CaIrSi$_3$ and plotted as $\lambda^{-2}\left(T\right)/\lambda^{-2}\left(0\right)$ versus the reduced temperature $T/T{\mathrm{c}}$ are shown in Fig.~\ref{fig7:lvt}. The $\lambda(T)$ curves can then be fit within the local (London) approximation~\cite{Tinkham} for an $s$-wave gap superconductor in the clean limit using the following expression:

\begin{equation}
\frac{\lambda^{-2}(T)}{\lambda^{-2}(0)}=1+2\int_{\Delta(T)}^{\infty}\left(\frac{\partial f}{\partial E}\right)\frac{E}{\sqrt{E^2-\Delta^2(T)}}dE,
\end{equation}
where $f=[1+\exp(E/k_{\mathrm{B}}T)]^{-1}$ is the Fermi function. The temperature dependence of the gap is approximated by $\Delta(T)=\Delta(0){\tanh}[1.82(1.018(T_{\mathrm{C}}/T-1))^{0.51}]$~\cite{Carrington03}.
As can be seen from Fig.~\ref{fig7:lvt} the temperature dependence of $\sigma_{\mathrm{sc}}$ for both samples is very well described by this isotropic $s$-wave model giving $\Delta(0)=0.81(1)$ and $\Delta(0)=0.38(1)$~meV, and BCS ratios $2\Delta\left(0\right)/k_BT_c$ of 3.8(2) and 5.4(2), for CaPtSi$_3$ and CaIrSi$_3$ respectively. The value for the BCS ratio is CaPtSi$_3$ is slightly higher that the 3.5 expected in the weak-coupling limit. The higher value obtained for CaIrSi$_3$, along with the strong field dependence of $\sigma_{\mathrm{sc}}$ shown in Fig.~\ref{fig5:svb}, could be evidence of a strong-coupling and/or multigap behavior with each gap having a similar temperature dependence. Such a suggestion is consistent with the departure from a pure $s$-wave behavior seen in the specific heat of CaIrSi$_3$, although the specific heat results for CaIrSi$_3$ are generally well explained by a weak-coupling BCS theory~\cite{Eguchi11}.

\begin{table}
\caption{Superconducting parameters of noncentrosymmetric  CaPtSi$_3$ and CaIrSi$_3$ determined from the dc magnetization and muon spectroscopy data.}
\label{table_of_SC_parameters}
\begin{center}
\begin{tabular}[b]{lll}\hline\hline
{}~~~~~&CaPtSi$_3$~~~~~&CaIrSi$_3$\\\hline
$T_c$,~(K)~& 2.30(5)& 3.50(5)  \\
$\lambda(0)$,~(nm)~& 448(6)& 150(7)  \\
$\Delta(0)$,~(meV)  & 0.38(1)&  0.81(1) \\
BCS ratio & 3.8(2)& 5.4(2)  \\\hline\hline
\end{tabular}
\par\medskip\footnotesize
\end{center}
\end{table} 

\section{Summary}
In summary, we have investigated the superconducting compounds CaPtSi$_3$ and CaIrSi$_3$ by using muon-spin relaxation and rotation. There is no evidence of time-reversal symmetry breaking in either material, at least within the sensitivity of $\mu$SR. The  superconducting parameters determined from this study are summarized in Table~\ref{table_of_SC_parameters}. The temperature dependence of the penetration depths for both materials are consistent with $s$-wave isotropic gaps. The BCS ratios place both materials in the intermediate to strong-coupling limit.

\begin{acknowledgments}
We acknowledge the EPSRC, UK for providing funding (grant number EP/I007210/1). We thank T. E. Orton for technical support. Some of the equipment used in this research at the University of Warwick was obtained through the Science City Advanced Materials: Creating and Characterizing Next Generation Advanced Materials Project, with support from Advantage West Midlands (AWM) and partly funded by the European Regional Development Fund (ERDF). 
\end{acknowledgments}

\bibliography{NCS}

\begin{thebibliography}{43}
\expandafter\ifx\csname natexlab\endcsname\relax\def\natexlab#1{#1}\fi
\expandafter\ifx\csname bibnamefont\endcsname\relax
  \def\bibnamefont#1{#1}\fi
\expandafter\ifx\csname bibfnamefont\endcsname\relax
  \def\bibfnamefont#1{#1}\fi
\expandafter\ifx\csname citenamefont\endcsname\relax
  \def\citenamefont#1{#1}\fi
\expandafter\ifx\csname url\endcsname\relax
  \def\url#1{\texttt{#1}}\fi
\expandafter\ifx\csname urlprefix\endcsname\relax\def\urlprefix{URL }\fi
\providecommand{\bibinfo}[2]{#2}
\providecommand{\eprint}[2][]{\url{#2}}

\bibitem[{\citenamefont{Bauer et~al.}(2004)\citenamefont{Bauer, Hilscher,
  Michor, Paul, Scheidt, Gribanov, Seropegin, No\"{e}l, Sigrist, and
  Rogl}}]{Bauer04}
\bibinfo{author}{\bibfnamefont{E.}~\bibnamefont{Bauer}},
  \bibinfo{author}{\bibfnamefont{G.}~\bibnamefont{Hilscher}},
  \bibinfo{author}{\bibfnamefont{H.}~\bibnamefont{Michor}},
  \bibinfo{author}{\bibfnamefont{{\relax Ch}.}~\bibnamefont{Paul}},
  \bibinfo{author}{\bibfnamefont{E.~W.} \bibnamefont{Scheidt}},
  \bibinfo{author}{\bibfnamefont{A.}~\bibnamefont{Gribanov}},
  \bibinfo{author}{\bibfnamefont{{\relax Yu}.}~\bibnamefont{Seropegin}},
  \bibinfo{author}{\bibfnamefont{H.}~\bibnamefont{No\"{e}l}},
  \bibinfo{author}{\bibfnamefont{M.}~\bibnamefont{Sigrist}}, \bibnamefont{and}
  \bibinfo{author}{\bibfnamefont{P.}~\bibnamefont{Rogl}},
  \bibinfo{journal}{Phys. Rev. Lett.} \textbf{\bibinfo{volume}{92}},
  \bibinfo{pages}{027003} (\bibinfo{year}{2004}).

\bibitem[{\citenamefont{Gor'kov and Rashba}(2001)}]{Gorkov01}
\bibinfo{author}{\bibfnamefont{L.~P.} \bibnamefont{Gor'kov}} \bibnamefont{and}
  \bibinfo{author}{\bibfnamefont{E.~I.} \bibnamefont{Rashba}},
  \bibinfo{journal}{Phys. Rev. Lett.} \textbf{\bibinfo{volume}{87}},
  \bibinfo{pages}{037004} (\bibinfo{year}{2001}).

\bibitem[{\citenamefont{Frigeri et~al.}(2004)\citenamefont{Frigeri, Agterberg,
  Koga, and Sigrist}}]{Frigeri04}
\bibinfo{author}{\bibfnamefont{P.~A.} \bibnamefont{Frigeri}},
  \bibinfo{author}{\bibfnamefont{D.~F.} \bibnamefont{Agterberg}},
  \bibinfo{author}{\bibfnamefont{A.}~\bibnamefont{Koga}}, \bibnamefont{and}
  \bibinfo{author}{\bibfnamefont{M.}~\bibnamefont{Sigrist}},
  \bibinfo{journal}{Phys. Rev. Lett.} \textbf{\bibinfo{volume}{92}},
  \bibinfo{pages}{097001} (\bibinfo{year}{2004}).

\bibitem[{\citenamefont{Frigeri et~al.}(2006)\citenamefont{Frigeri, Agterberg,
  Milat, and Sigrist}}]{Frigeri06}
\bibinfo{author}{\bibfnamefont{P.~A.} \bibnamefont{Frigeri}},
  \bibinfo{author}{\bibfnamefont{D.~F.} \bibnamefont{Agterberg}},
  \bibinfo{author}{\bibfnamefont{I.}~\bibnamefont{Milat}}, \bibnamefont{and}
  \bibinfo{author}{\bibfnamefont{M.}~\bibnamefont{Sigrist}},
  \bibinfo{journal}{Eur. Phys J. B} \textbf{\bibinfo{volume}{54}},
  \bibinfo{pages}{435} (\bibinfo{year}{2006}).

\bibitem[{\citenamefont{Sigrist}(2009)}]{Sigrist08}
\bibinfo{author}{\bibfnamefont{M.}~\bibnamefont{Sigrist}},
  \bibinfo{journal}{AIP Conf. Proc} \textbf{\bibinfo{volume}{1162}},
  \bibinfo{pages}{55} (\bibinfo{year}{2009}).

\bibitem[{\citenamefont{Bauer and Sigrist}(2012)}]{Bauer12}
\bibinfo{author}{\bibfnamefont{E.}~\bibnamefont{Bauer}} \bibnamefont{and}
  \bibinfo{author}{\bibfnamefont{M.}~\bibnamefont{Sigrist}},
  \emph{\bibinfo{title}{Non-Centrosymmetric Superconductors: Introduction and
  Overview, Lecture Notes in Physics}} (\bibinfo{publisher}{Springer-Verlag
  Berlin Heidelberg}, \bibinfo{year}{2012}).

\bibitem[{\citenamefont{Schenck}(1985)}]{Schenck}
\bibinfo{author}{\bibfnamefont{A.}~\bibnamefont{Schenck}},
  \emph{\bibinfo{title}{Muon Spin Rotation Spectroscopy Principles and
  Applications in Solid State Physics}} (\bibinfo{publisher}{Taylor and
  Francis}, \bibinfo{year}{1985}).

\bibitem[{\citenamefont{Lee et~al.}(1998)\citenamefont{Lee, Kilcoyne, and
  Cywinski}}]{Lee98}
\bibinfo{author}{\bibfnamefont{S.~L.} \bibnamefont{Lee}},
  \bibinfo{author}{\bibfnamefont{S.~H.} \bibnamefont{Kilcoyne}},
  \bibnamefont{and} \bibinfo{author}{\bibfnamefont{R.}~\bibnamefont{Cywinski}},
  \emph{\bibinfo{title}{Muon Science: Muons in Physics, Chemistry and
  Materials}} (\bibinfo{publisher}{SUSSP and IOP}, \bibinfo{year}{1998}).

\bibitem[{\citenamefont{Sonier}(2007)}]{Sonier07}
\bibinfo{author}{\bibfnamefont{J.~E.} \bibnamefont{Sonier}},
  \bibinfo{journal}{Rep. Prog. Phys.} \textbf{\bibinfo{volume}{70}},
  \bibinfo{pages}{1717} (\bibinfo{year}{2007}).

\bibitem[{\citenamefont{Yaouanc and de~R\'{e}otier}(2011)}]{Yaouanc}
\bibinfo{author}{\bibfnamefont{A.}~\bibnamefont{Yaouanc}} \bibnamefont{and}
  \bibinfo{author}{\bibfnamefont{P.~D.} \bibnamefont{de~R\'{e}otier}},
  \emph{\bibinfo{title}{Muon Spin Rotation, Relaxation, and Resonance}}
  (\bibinfo{publisher}{Oxford University Press}, \bibinfo{year}{2011}).

\bibitem[{\citenamefont{Luke et~al.}(1998)\citenamefont{Luke, Fudamoto, Kojima,
  Larkin, Merrin, Nachumi, Uemura, Maeno, Mao, Mori et~al.}}]{Luke98}
\bibinfo{author}{\bibfnamefont{G.~M.} \bibnamefont{Luke}},
  \bibinfo{author}{\bibfnamefont{Y.}~\bibnamefont{Fudamoto}},
  \bibinfo{author}{\bibfnamefont{K.~M.} \bibnamefont{Kojima}},
  \bibinfo{author}{\bibfnamefont{M.~I.} \bibnamefont{Larkin}},
  \bibinfo{author}{\bibfnamefont{J.}~\bibnamefont{Merrin}},
  \bibinfo{author}{\bibfnamefont{B.}~\bibnamefont{Nachumi}},
  \bibinfo{author}{\bibfnamefont{Y.~J.} \bibnamefont{Uemura}},
  \bibinfo{author}{\bibfnamefont{Y.}~\bibnamefont{Maeno}},
  \bibinfo{author}{\bibfnamefont{Z.~Q.} \bibnamefont{Mao}},
  \bibinfo{author}{\bibfnamefont{Y.}~\bibnamefont{Mori}}, \bibnamefont{et~al.},
  \bibinfo{journal}{Nature} \textbf{\bibinfo{volume}{394}},
  \bibinfo{pages}{558} (\bibinfo{year}{1998}).

\bibitem[{\citenamefont{Luke et~al.}(1993)\citenamefont{Luke, Keren, Le, Wu,
  Uemura, Bonn, Taillefer, and Garrett}}]{Luke93}
\bibinfo{author}{\bibfnamefont{G.~M.} \bibnamefont{Luke}},
  \bibinfo{author}{\bibfnamefont{A.}~\bibnamefont{Keren}},
  \bibinfo{author}{\bibfnamefont{L.~P.} \bibnamefont{Le}},
  \bibinfo{author}{\bibfnamefont{W.~D.} \bibnamefont{Wu}},
  \bibinfo{author}{\bibfnamefont{Y.~J.} \bibnamefont{Uemura}},
  \bibinfo{author}{\bibfnamefont{D.~A.} \bibnamefont{Bonn}},
  \bibinfo{author}{\bibfnamefont{L.}~\bibnamefont{Taillefer}},
  \bibnamefont{and} \bibinfo{author}{\bibfnamefont{J.~D.}
  \bibnamefont{Garrett}}, \bibinfo{journal}{Phys. Rev. Lett.}
  \textbf{\bibinfo{volume}{71}}, \bibinfo{pages}{1466} (\bibinfo{year}{1993}).

\bibitem[{\citenamefont{de~R\'{e}otier
  et~al.}(1995)\citenamefont{de~R\'{e}otier, Huxley, Yaouanc, Flouquet,
  Bonville, Impert, Pari, Gubbens, and Mulders}}]{deReotier95}
\bibinfo{author}{\bibfnamefont{P.~D.} \bibnamefont{de~R\'{e}otier}},
  \bibinfo{author}{\bibfnamefont{A.}~\bibnamefont{Huxley}},
  \bibinfo{author}{\bibfnamefont{A.}~\bibnamefont{Yaouanc}},
  \bibinfo{author}{\bibfnamefont{J.}~\bibnamefont{Flouquet}},
  \bibinfo{author}{\bibfnamefont{P.}~\bibnamefont{Bonville}},
  \bibinfo{author}{\bibfnamefont{P.}~\bibnamefont{Impert}},
  \bibinfo{author}{\bibfnamefont{P.}~\bibnamefont{Pari}},
  \bibinfo{author}{\bibfnamefont{P.~C.~M.} \bibnamefont{Gubbens}},
  \bibnamefont{and} \bibinfo{author}{\bibfnamefont{A.~M.}
  \bibnamefont{Mulders}}, \bibinfo{journal}{Phys. Lett. A}
  \textbf{\bibinfo{volume}{205}}, \bibinfo{pages}{239} (\bibinfo{year}{1995}).

\bibitem[{\citenamefont{Heffner et~al.}(1990)\citenamefont{Heffner, Smith,
  Willis, Birrer, Baines, Gygax, Hitti, Lippelt, Ott, Schenck
  et~al.}}]{Heffner90}
\bibinfo{author}{\bibfnamefont{R.~H.} \bibnamefont{Heffner}},
  \bibinfo{author}{\bibfnamefont{J.~L.} \bibnamefont{Smith}},
  \bibinfo{author}{\bibfnamefont{J.~O.} \bibnamefont{Willis}},
  \bibinfo{author}{\bibfnamefont{P.}~\bibnamefont{Birrer}},
  \bibinfo{author}{\bibfnamefont{C.}~\bibnamefont{Baines}},
  \bibinfo{author}{\bibfnamefont{F.~N.} \bibnamefont{Gygax}},
  \bibinfo{author}{\bibfnamefont{B.}~\bibnamefont{Hitti}},
  \bibinfo{author}{\bibfnamefont{E.}~\bibnamefont{Lippelt}},
  \bibinfo{author}{\bibfnamefont{H.~R.} \bibnamefont{Ott}},
  \bibinfo{author}{\bibfnamefont{A.}~\bibnamefont{Schenck}},
  \bibnamefont{et~al.}, \bibinfo{journal}{Phys. Rev. Lett.}
  \textbf{\bibinfo{volume}{65}}, \bibinfo{pages}{2816} (\bibinfo{year}{1990}).

\bibitem[{\citenamefont{Aoki et~al.}(2003)\citenamefont{Aoki, Tsuchiya,
  Kanayama, Saha, Sugawara, Sato, Higemoto, Koda, Ohishi, Nishiyama
  et~al.}}]{Aoki03}
\bibinfo{author}{\bibfnamefont{Y.}~\bibnamefont{Aoki}},
  \bibinfo{author}{\bibfnamefont{A.}~\bibnamefont{Tsuchiya}},
  \bibinfo{author}{\bibfnamefont{T.}~\bibnamefont{Kanayama}},
  \bibinfo{author}{\bibfnamefont{S.~R.} \bibnamefont{Saha}},
  \bibinfo{author}{\bibfnamefont{H.}~\bibnamefont{Sugawara}},
  \bibinfo{author}{\bibfnamefont{H.}~\bibnamefont{Sato}},
  \bibinfo{author}{\bibfnamefont{W.}~\bibnamefont{Higemoto}},
  \bibinfo{author}{\bibfnamefont{A.}~\bibnamefont{Koda}},
  \bibinfo{author}{\bibfnamefont{K.}~\bibnamefont{Ohishi}},
  \bibinfo{author}{\bibfnamefont{K.}~\bibnamefont{Nishiyama}},
  \bibnamefont{et~al.}, \bibinfo{journal}{Phys. Rev. Lett.}
  \textbf{\bibinfo{volume}{91}}, \bibinfo{pages}{067003}
  (\bibinfo{year}{2003}).

\bibitem[{\citenamefont{Hillier et~al.}(2009)\citenamefont{Hillier,
  Quintanilla, and Cywinski}}]{Hillier09}
\bibinfo{author}{\bibfnamefont{A.~D.} \bibnamefont{Hillier}},
  \bibinfo{author}{\bibfnamefont{J.}~\bibnamefont{Quintanilla}},
  \bibnamefont{and} \bibinfo{author}{\bibfnamefont{R.}~\bibnamefont{Cywinski}},
  \bibinfo{journal}{Phys. Rev. Lett.} \textbf{\bibinfo{volume}{102}},
  \bibinfo{pages}{117007} (\bibinfo{year}{2009}).

\bibitem[{\citenamefont{Maisuradze et~al.}(2010)\citenamefont{Maisuradze,
  Schnelle, Khasanov, Gumeniuk, Nicklas, Rosner, Leithe-Jasper, Grin, Amato,
  and Thalmeier}}]{Maisuradze10}
\bibinfo{author}{\bibfnamefont{A.}~\bibnamefont{Maisuradze}},
  \bibinfo{author}{\bibfnamefont{W.}~\bibnamefont{Schnelle}},
  \bibinfo{author}{\bibfnamefont{R.}~\bibnamefont{Khasanov}},
  \bibinfo{author}{\bibfnamefont{R.}~\bibnamefont{Gumeniuk}},
  \bibinfo{author}{\bibfnamefont{M.}~\bibnamefont{Nicklas}},
  \bibinfo{author}{\bibfnamefont{H.}~\bibnamefont{Rosner}},
  \bibinfo{author}{\bibfnamefont{A.}~\bibnamefont{Leithe-Jasper}},
  \bibinfo{author}{\bibfnamefont{{\relax Yu}.}~\bibnamefont{Grin}},
  \bibinfo{author}{\bibfnamefont{A.}~\bibnamefont{Amato}}, \bibnamefont{and}
  \bibinfo{author}{\bibfnamefont{P.}~\bibnamefont{Thalmeier}},
  \bibinfo{journal}{Phys. Rev. B} \textbf{\bibinfo{volume}{82}},
  \bibinfo{pages}{024524} (\bibinfo{year}{2010}).

\bibitem[{\citenamefont{Shu et~al.}(2011)\citenamefont{Shu, Higemoto, Aoki,
  Hillier, Ohishi, Ishida, Kadono, Koda, Bernal, MacLaughlin et~al.}}]{Shu11}
\bibinfo{author}{\bibfnamefont{L.}~\bibnamefont{Shu}},
  \bibinfo{author}{\bibfnamefont{W.}~\bibnamefont{Higemoto}},
  \bibinfo{author}{\bibfnamefont{Y.}~\bibnamefont{Aoki}},
  \bibinfo{author}{\bibfnamefont{A.~D.} \bibnamefont{Hillier}},
  \bibinfo{author}{\bibfnamefont{K.}~\bibnamefont{Ohishi}},
  \bibinfo{author}{\bibfnamefont{K.}~\bibnamefont{Ishida}},
  \bibinfo{author}{\bibfnamefont{R.}~\bibnamefont{Kadono}},
  \bibinfo{author}{\bibfnamefont{A.}~\bibnamefont{Koda}},
  \bibinfo{author}{\bibfnamefont{O.~O.} \bibnamefont{Bernal}},
  \bibinfo{author}{\bibfnamefont{D.~E.} \bibnamefont{MacLaughlin}},
  \bibnamefont{et~al.}, \bibinfo{journal}{Phys. Rev. B}
  \textbf{\bibinfo{volume}{83}}, \bibinfo{pages}{100504}
  (\bibinfo{year}{2011}).

\bibitem[{\citenamefont{Hillier et~al.}(2012)\citenamefont{Hillier,
  Quintanilla, and Cywinski}}]{Hillier12}
\bibinfo{author}{\bibfnamefont{A.~D.} \bibnamefont{Hillier}},
  \bibinfo{author}{\bibfnamefont{J.}~\bibnamefont{Quintanilla}},
  \bibnamefont{and} \bibinfo{author}{\bibfnamefont{R.}~\bibnamefont{Cywinski}},
  \bibinfo{journal}{Phys. Rev. Lett.} \textbf{\bibinfo{volume}{109}},
  \bibinfo{pages}{097001} (\bibinfo{year}{2012}).

\bibitem[{\citenamefont{Biswas et~al.}(2013)\citenamefont{Biswas, Luetkens,
  Neupert, St\"{u}rzer, Baines, Pascua, Schnyder, Fischer, Goryo, Lees
  et~al.}}]{Biswas13}
\bibinfo{author}{\bibfnamefont{P.~K.} \bibnamefont{Biswas}},
  \bibinfo{author}{\bibfnamefont{H.}~\bibnamefont{Luetkens}},
  \bibinfo{author}{\bibfnamefont{T.}~\bibnamefont{Neupert}},
  \bibinfo{author}{\bibfnamefont{T.}~\bibnamefont{St\"{u}rzer}},
  \bibinfo{author}{\bibfnamefont{C.}~\bibnamefont{Baines}},
  \bibinfo{author}{\bibfnamefont{G.}~\bibnamefont{Pascua}},
  \bibinfo{author}{\bibfnamefont{A.~P.} \bibnamefont{Schnyder}},
  \bibinfo{author}{\bibfnamefont{M.~H.} \bibnamefont{Fischer}},
  \bibinfo{author}{\bibfnamefont{J.}~\bibnamefont{Goryo}},
  \bibinfo{author}{\bibfnamefont{M.~R.} \bibnamefont{Lees}},
  \bibnamefont{et~al.}, \bibinfo{journal}{Phys. Rev. B}
  \textbf{\bibinfo{volume}{87}}, \bibinfo{pages}{180503}
  (\bibinfo{year}{2013}).

\bibitem[{\citenamefont{Singh et~al.}(2014)\citenamefont{Singh, Hillier,
  Mazidian, Quintanilla, Annett, Paul, Balakrishnan, and Lees}}]{Singh14}
\bibinfo{author}{\bibfnamefont{R.~P.} \bibnamefont{Singh}},
  \bibinfo{author}{\bibfnamefont{A.}~\bibnamefont{Hillier}},
  \bibinfo{author}{\bibfnamefont{B.}~\bibnamefont{Mazidian}},
  \bibinfo{author}{\bibfnamefont{J.}~\bibnamefont{Quintanilla}},
  \bibinfo{author}{\bibfnamefont{J.}~\bibnamefont{Annett}},
  \bibinfo{author}{\bibfnamefont{D.~M.} \bibnamefont{Paul}},
  \bibinfo{author}{\bibfnamefont{G.}~\bibnamefont{Balakrishnan}},
  \bibnamefont{and} \bibinfo{author}{\bibfnamefont{M.~R.} \bibnamefont{Lees}},
  \bibinfo{journal}{Phys. Rev. Lett.} \textbf{\bibinfo{volume}{112}},
  \bibinfo{pages}{107002} (\bibinfo{year}{2014}).

\bibitem[{\citenamefont{Quintanilla et~al.}(2010)\citenamefont{Quintanilla,
  Hillier, Annett, and Cywinski}}]{Quintanilla10}
\bibinfo{author}{\bibfnamefont{J.}~\bibnamefont{Quintanilla}},
  \bibinfo{author}{\bibfnamefont{A.~D.} \bibnamefont{Hillier}},
  \bibinfo{author}{\bibfnamefont{J.~F.} \bibnamefont{Annett}},
  \bibnamefont{and} \bibinfo{author}{\bibfnamefont{R.}~\bibnamefont{Cywinski}},
  \bibinfo{journal}{Phys. Rev. B} \textbf{\bibinfo{volume}{82}},
  \bibinfo{pages}{174511} (\bibinfo{year}{2010}).

\bibitem[{\citenamefont{Karki et~al.}(2010)\citenamefont{Karki, Xiong, Vekhter,
  Browne, Adams, Young, Thomas, Chan, Kim, and Prozorov}}]{Karki201a}
\bibinfo{author}{\bibfnamefont{A.~B.} \bibnamefont{Karki}},
  \bibinfo{author}{\bibfnamefont{Y.~M.} \bibnamefont{Xiong}},
  \bibinfo{author}{\bibfnamefont{I.}~\bibnamefont{Vekhter}},
  \bibinfo{author}{\bibfnamefont{D.}~\bibnamefont{Browne}},
  \bibinfo{author}{\bibfnamefont{P.~W.} \bibnamefont{Adams}},
  \bibinfo{author}{\bibfnamefont{D.~P.} \bibnamefont{Young}},
  \bibinfo{author}{\bibfnamefont{K.~R.} \bibnamefont{Thomas}},
  \bibinfo{author}{\bibfnamefont{J.~Y.} \bibnamefont{Chan}},
  \bibinfo{author}{\bibfnamefont{H.}~\bibnamefont{Kim}}, \bibnamefont{and}
  \bibinfo{author}{\bibfnamefont{R.}~\bibnamefont{Prozorov}},
  \bibinfo{journal}{Phys. Rev. B} \textbf{\bibinfo{volume}{82}},
  \bibinfo{pages}{064512} (\bibinfo{year}{2010}).

\bibitem[{\citenamefont{Anand et~al.}(2011)\citenamefont{Anand, Hillier,
  Adroja, Strydom, Michor, McEwen, and Rainford}}]{Anand11}
\bibinfo{author}{\bibfnamefont{V.}~\bibnamefont{Anand}},
  \bibinfo{author}{\bibfnamefont{A.~D.} \bibnamefont{Hillier}},
  \bibinfo{author}{\bibfnamefont{D.~T.} \bibnamefont{Adroja}},
  \bibinfo{author}{\bibfnamefont{A.~M.} \bibnamefont{Strydom}},
  \bibinfo{author}{\bibfnamefont{H.}~\bibnamefont{Michor}},
  \bibinfo{author}{\bibfnamefont{K.~A.} \bibnamefont{McEwen}},
  \bibnamefont{and} \bibinfo{author}{\bibfnamefont{B.~D.}
  \bibnamefont{Rainford}}, \bibinfo{journal}{Phys. Rev. B}
  \textbf{\bibinfo{volume}{83}}, \bibinfo{pages}{064552}
  (\bibinfo{year}{2011}).

\bibitem[{\citenamefont{Klimczuk et~al.}(2007)\citenamefont{Klimczuk, Ronning,
  Sidorov, Cava, and Thompson}}]{Klimczuk2007a}
\bibinfo{author}{\bibfnamefont{T.}~\bibnamefont{Klimczuk}},
  \bibinfo{author}{\bibfnamefont{F.}~\bibnamefont{Ronning}},
  \bibinfo{author}{\bibfnamefont{V.}~\bibnamefont{Sidorov}},
  \bibinfo{author}{\bibfnamefont{R.~J.} \bibnamefont{Cava}}, \bibnamefont{and}
  \bibinfo{author}{\bibfnamefont{J.~D.} \bibnamefont{Thompson}},
  \bibinfo{journal}{Phys. Rev. Lett.} \textbf{\bibinfo{volume}{99}},
  \bibinfo{pages}{257004} (\bibinfo{year}{2007}).

\bibitem[{\citenamefont{Karki et~al.}(2011)\citenamefont{Karki, Xiong,
  Haldolaarachchige, Stadler, Vekhter, Adams, Young, Phelan, and
  Chan}}]{Karki2011a}
\bibinfo{author}{\bibfnamefont{A.~B.} \bibnamefont{Karki}},
  \bibinfo{author}{\bibfnamefont{Y.~M.} \bibnamefont{Xiong}},
  \bibinfo{author}{\bibfnamefont{N.}~\bibnamefont{Haldolaarachchige}},
  \bibinfo{author}{\bibfnamefont{S.}~\bibnamefont{Stadler}},
  \bibinfo{author}{\bibfnamefont{I.}~\bibnamefont{Vekhter}},
  \bibinfo{author}{\bibfnamefont{P.~W.} \bibnamefont{Adams}},
  \bibinfo{author}{\bibfnamefont{D.~P.} \bibnamefont{Young}},
  \bibinfo{author}{\bibfnamefont{W.~A.} \bibnamefont{Phelan}},
  \bibnamefont{and} \bibinfo{author}{\bibfnamefont{J.~Y.} \bibnamefont{Chan}},
  \bibinfo{journal}{Phys. Rev. B} \textbf{\bibinfo{volume}{83}},
  \bibinfo{pages}{144525} (\bibinfo{year}{2011}).

\bibitem[{\citenamefont{Biswas et~al.}(2011)\citenamefont{Biswas, Lees,
  Hillier, Smith, Marshall, and Paul}}]{Biswas11}
\bibinfo{author}{\bibfnamefont{P.~K.} \bibnamefont{Biswas}},
  \bibinfo{author}{\bibfnamefont{M.~R.} \bibnamefont{Lees}},
  \bibinfo{author}{\bibfnamefont{A.~D.} \bibnamefont{Hillier}},
  \bibinfo{author}{\bibfnamefont{R.~I.} \bibnamefont{Smith}},
  \bibinfo{author}{\bibfnamefont{W.~G.} \bibnamefont{Marshall}},
  \bibnamefont{and} \bibinfo{author}{\bibfnamefont{D.~M.} \bibnamefont{Paul}},
  \bibinfo{journal}{Phys. Rev. B} \textbf{\bibinfo{volume}{84}},
  \bibinfo{pages}{184529} (\bibinfo{year}{2011}).

\bibitem[{\citenamefont{Smidman et~al.}(2014)\citenamefont{Smidman, Hillier,
  Adroja, Lees, Anand, Singh, Smith, Paul, and Balakrishnan}}]{Smidman14}
\bibinfo{author}{\bibfnamefont{M.}~\bibnamefont{Smidman}},
  \bibinfo{author}{\bibfnamefont{A.~D.} \bibnamefont{Hillier}},
  \bibinfo{author}{\bibfnamefont{D.~T.} \bibnamefont{Adroja}},
  \bibinfo{author}{\bibfnamefont{M.~R.} \bibnamefont{Lees}},
  \bibinfo{author}{\bibfnamefont{V.~K.} \bibnamefont{Anand}},
  \bibinfo{author}{\bibfnamefont{R.~P.} \bibnamefont{Singh}},
  \bibinfo{author}{\bibfnamefont{R.~I.} \bibnamefont{Smith}},
  \bibinfo{author}{\bibfnamefont{D.~M.} \bibnamefont{Paul}}, \bibnamefont{and}
  \bibinfo{author}{\bibfnamefont{G.}~\bibnamefont{Balakrishnan}},
  \bibinfo{journal}{Phys. Rev. B} \textbf{\bibinfo{volume}{89}},
  \bibinfo{pages}{094509} (\bibinfo{year}{2014}).

\bibitem[{\citenamefont{Haen et~al.}(1985)\citenamefont{Haen, Lejay, Chevalier,
  Lloret, Etourneau, and Sera}}]{Haen85}
\bibinfo{author}{\bibfnamefont{P.}~\bibnamefont{Haen}},
  \bibinfo{author}{\bibfnamefont{P.}~\bibnamefont{Lejay}},
  \bibinfo{author}{\bibfnamefont{B.}~\bibnamefont{Chevalier}},
  \bibinfo{author}{\bibfnamefont{B.}~\bibnamefont{Lloret}},
  \bibinfo{author}{\bibfnamefont{J.}~\bibnamefont{Etourneau}},
  \bibnamefont{and} \bibinfo{author}{\bibfnamefont{M.}~\bibnamefont{Sera}},
  \bibinfo{journal}{J. Less-Common Met.} \textbf{\bibinfo{volume}{110}},
  \bibinfo{pages}{321} (\bibinfo{year}{1985}).

\bibitem[{\citenamefont{Iwamoto et~al.}(1995)\citenamefont{Iwamoto, Ueda,
  Kohara, and Yamada}}]{Iwamoto95}
\bibinfo{author}{\bibfnamefont{Y.}~\bibnamefont{Iwamoto}},
  \bibinfo{author}{\bibfnamefont{K.}~\bibnamefont{Ueda}},
  \bibinfo{author}{\bibfnamefont{T.}~\bibnamefont{Kohara}}, \bibnamefont{and}
  \bibinfo{author}{\bibfnamefont{Y.}~\bibnamefont{Yamada}},
  \bibinfo{journal}{Physica B} \textbf{\bibinfo{volume}{206-207}},
  \bibinfo{pages}{276} (\bibinfo{year}{1995}).

\bibitem[{\citenamefont{Kimura et~al.}(2005)\citenamefont{Kimura, Ito, Saitoh,
  Umeda, Aoki, and Terashima}}]{Kimura05}
\bibinfo{author}{\bibfnamefont{N.}~\bibnamefont{Kimura}},
  \bibinfo{author}{\bibfnamefont{K.}~\bibnamefont{Ito}},
  \bibinfo{author}{\bibfnamefont{K.}~\bibnamefont{Saitoh}},
  \bibinfo{author}{\bibfnamefont{Y.}~\bibnamefont{Umeda}},
  \bibinfo{author}{\bibfnamefont{H.}~\bibnamefont{Aoki}}, \bibnamefont{and}
  \bibinfo{author}{\bibfnamefont{T.}~\bibnamefont{Terashima}},
  \bibinfo{journal}{Phys. Rev. Lett.} \textbf{\bibinfo{volume}{95}},
  \bibinfo{pages}{247004} (\bibinfo{year}{2005}).

\bibitem[{\citenamefont{Kimura et~al.}(2007)\citenamefont{Kimura, Ito, Aoki,
  Uji, and Terashima}}]{Kimura07}
\bibinfo{author}{\bibfnamefont{N.}~\bibnamefont{Kimura}},
  \bibinfo{author}{\bibfnamefont{K.}~\bibnamefont{Ito}},
  \bibinfo{author}{\bibfnamefont{H.}~\bibnamefont{Aoki}},
  \bibinfo{author}{\bibfnamefont{S.}~\bibnamefont{Uji}}, \bibnamefont{and}
  \bibinfo{author}{\bibfnamefont{T.}~\bibnamefont{Terashima}},
  \bibinfo{journal}{Phys. Rev. Lett.} \textbf{\bibinfo{volume}{98}},
  \bibinfo{pages}{197001} (\bibinfo{year}{2007}).

\bibitem[{\citenamefont{Sugitani et~al.}(2006)\citenamefont{Sugitani, Okuda,
  Shishido, Yamada, Thamizhavel, Yamamoto, D.~Matsuda, Haga, Takeuchi, Settai
  et~al.}}]{Sugitani06}
\bibinfo{author}{\bibfnamefont{I.}~\bibnamefont{Sugitani}},
  \bibinfo{author}{\bibfnamefont{Y.}~\bibnamefont{Okuda}},
  \bibinfo{author}{\bibfnamefont{H.}~\bibnamefont{Shishido}},
  \bibinfo{author}{\bibfnamefont{T.}~\bibnamefont{Yamada}},
  \bibinfo{author}{\bibfnamefont{A.}~\bibnamefont{Thamizhavel}},
  \bibinfo{author}{\bibfnamefont{E.}~\bibnamefont{Yamamoto}},
  \bibinfo{author}{\bibfnamefont{T.}~\bibnamefont{D.~Matsuda}},
  \bibinfo{author}{\bibfnamefont{Y.}~\bibnamefont{Haga}},
  \bibinfo{author}{\bibfnamefont{T.}~\bibnamefont{Takeuchi}},
  \bibinfo{author}{\bibfnamefont{R.}~\bibnamefont{Settai}},
  \bibnamefont{et~al.}, \bibinfo{journal}{J. Phys. Soc. Jpn.}
  \textbf{\bibinfo{volume}{75}}, \bibinfo{pages}{043703}
  (\bibinfo{year}{2006}).

\bibitem[{\citenamefont{Mukuda et~al.}(2008)\citenamefont{Mukuda, Fujii, Ohara,
  Harada, Yashima, Kitaoka, Okuda, Settai, and Onuki}}]{Mukuda08}
\bibinfo{author}{\bibfnamefont{H.}~\bibnamefont{Mukuda}},
  \bibinfo{author}{\bibfnamefont{T.}~\bibnamefont{Fujii}},
  \bibinfo{author}{\bibfnamefont{T.}~\bibnamefont{Ohara}},
  \bibinfo{author}{\bibfnamefont{A.}~\bibnamefont{Harada}},
  \bibinfo{author}{\bibfnamefont{M.}~\bibnamefont{Yashima}},
  \bibinfo{author}{\bibfnamefont{Y.}~\bibnamefont{Kitaoka}},
  \bibinfo{author}{\bibfnamefont{Y.}~\bibnamefont{Okuda}},
  \bibinfo{author}{\bibfnamefont{R.}~\bibnamefont{Settai}}, \bibnamefont{and}
  \bibinfo{author}{\bibfnamefont{Y.}~\bibnamefont{Onuki}},
  \bibinfo{journal}{Phys. Rev. Lett.} \textbf{\bibinfo{volume}{100}},
  \bibinfo{pages}{107003} (\bibinfo{year}{2008}).

\bibitem[{\citenamefont{Eguchi et~al.}(2011)\citenamefont{Eguchi, Peets,
  Kriener, Maeno, Nishibori, Kumazawa, Banno, Maki, and Sawa}}]{Eguchi11}
\bibinfo{author}{\bibfnamefont{G.}~\bibnamefont{Eguchi}},
  \bibinfo{author}{\bibfnamefont{D.~C.} \bibnamefont{Peets}},
  \bibinfo{author}{\bibfnamefont{M.}~\bibnamefont{Kriener}},
  \bibinfo{author}{\bibfnamefont{Y.}~\bibnamefont{Maeno}},
  \bibinfo{author}{\bibfnamefont{E.}~\bibnamefont{Nishibori}},
  \bibinfo{author}{\bibfnamefont{Y.}~\bibnamefont{Kumazawa}},
  \bibinfo{author}{\bibfnamefont{K.}~\bibnamefont{Banno}},
  \bibinfo{author}{\bibfnamefont{S.}~\bibnamefont{Maki}}, \bibnamefont{and}
  \bibinfo{author}{\bibfnamefont{H.}~\bibnamefont{Sawa}},
  \bibinfo{journal}{Phys. Rev. B} \textbf{\bibinfo{volume}{83}},
  \bibinfo{pages}{24512} (\bibinfo{year}{2011}).

\bibitem[{\citenamefont{Equchi et~al.}(2010)\citenamefont{Equchi, Peets,
  Kriener, Maki, Nishibori, Sawa, and Maeno}}]{Eguchi_sces}
\bibinfo{author}{\bibfnamefont{G.}~\bibnamefont{Equchi}},
  \bibinfo{author}{\bibfnamefont{D.}~\bibnamefont{Peets}},
  \bibinfo{author}{\bibfnamefont{M.}~\bibnamefont{Kriener}},
  \bibinfo{author}{\bibfnamefont{S.}~\bibnamefont{Maki}},
  \bibinfo{author}{\bibfnamefont{E.}~\bibnamefont{Nishibori}},
  \bibinfo{author}{\bibfnamefont{H.}~\bibnamefont{Sawa}}, \bibnamefont{and}
  \bibinfo{author}{\bibfnamefont{Y.}~\bibnamefont{Maeno}},
  \bibinfo{journal}{Physica C} \textbf{\bibinfo{volume}{470}},
  \bibinfo{pages}{S762} (\bibinfo{year}{2010}).

\bibitem[{\citenamefont{King et~al.}(2013)\citenamefont{King, de~Renzi,
  Cottrell, Hillier, and Cox}}]{King14}
\bibinfo{author}{\bibfnamefont{P.~J.~C.} \bibnamefont{King}},
  \bibinfo{author}{\bibfnamefont{R.}~\bibnamefont{de~Renzi}},
  \bibinfo{author}{\bibfnamefont{S.~P.} \bibnamefont{Cottrell}},
  \bibinfo{author}{\bibfnamefont{A.~D.} \bibnamefont{Hillier}},
  \bibnamefont{and} \bibinfo{author}{\bibfnamefont{S.~F.~J.}
  \bibnamefont{Cox}}, \bibinfo{journal}{Phys. Scripta}
  \textbf{\bibinfo{volume}{88}}, \bibinfo{pages}{068502}
  (\bibinfo{year}{2013}).

\bibitem[{\citenamefont{Maisuradze et~al.}(2009)\citenamefont{Maisuradze,
  Khasanov, Shengelaya, and Keller}}]{Maisuradze09}
\bibinfo{author}{\bibfnamefont{A.}~\bibnamefont{Maisuradze}},
  \bibinfo{author}{\bibfnamefont{R.}~\bibnamefont{Khasanov}},
  \bibinfo{author}{\bibfnamefont{A.}~\bibnamefont{Shengelaya}},
  \bibnamefont{and} \bibinfo{author}{\bibfnamefont{H.}~\bibnamefont{Keller}},
  \bibinfo{journal}{J Phys.-Condens. Mat.} \textbf{\bibinfo{volume}{21}},
  \bibinfo{pages}{075701} (\bibinfo{year}{2009}).

\bibitem[{\citenamefont{Weber et~al.}(1993)\citenamefont{Weber, Amato, Gygax,
  Schenck, Maletta, Duginov, Grebinnik, Lazarev, Olshevsky, Pomjakushin
  et~al.}}]{Weber}
\bibinfo{author}{\bibfnamefont{M.}~\bibnamefont{Weber}},
  \bibinfo{author}{\bibfnamefont{A.}~\bibnamefont{Amato}},
  \bibinfo{author}{\bibfnamefont{F.~N.} \bibnamefont{Gygax}},
  \bibinfo{author}{\bibfnamefont{A.}~\bibnamefont{Schenck}},
  \bibinfo{author}{\bibfnamefont{H.}~\bibnamefont{Maletta}},
  \bibinfo{author}{\bibfnamefont{V.~N.} \bibnamefont{Duginov}},
  \bibinfo{author}{\bibfnamefont{V.~G.} \bibnamefont{Grebinnik}},
  \bibinfo{author}{\bibfnamefont{A.~B.} \bibnamefont{Lazarev}},
  \bibinfo{author}{\bibfnamefont{V.~G.} \bibnamefont{Olshevsky}},
  \bibinfo{author}{\bibfnamefont{V.~Y.} \bibnamefont{Pomjakushin}},
  \bibnamefont{et~al.}, \bibinfo{journal}{Phys. Rev. B}
  \textbf{\bibinfo{volume}{48}}, \bibinfo{pages}{13022} (\bibinfo{year}{1993}).

\bibitem[{\citenamefont{Brandt}(1988)}]{Brandt88}
\bibinfo{author}{\bibfnamefont{E.~H.} \bibnamefont{Brandt}},
  \bibinfo{journal}{J. Low Temp. Phys.} \textbf{\bibinfo{volume}{73}},
  \bibinfo{pages}{355} (\bibinfo{year}{1988}).

\bibitem[{\citenamefont{Brandt}(2003)}]{Brandt03}
\bibinfo{author}{\bibfnamefont{E.~H.} \bibnamefont{Brandt}},
  \bibinfo{journal}{Phys. Rev. B} \textbf{\bibinfo{volume}{68}},
  \bibinfo{pages}{054506} (\bibinfo{year}{2003}).

\bibitem[{\citenamefont{Tinkham}(1996)}]{Tinkham}
\bibinfo{author}{\bibfnamefont{M.}~\bibnamefont{Tinkham}},
  \emph{\bibinfo{title}{Introduction to Superconductivity}}
  (\bibinfo{publisher}{McGraw-Hill, New York}, \bibinfo{year}{1996}),
  \bibinfo{edition}{2nd} ed.

\bibitem[{\citenamefont{Carrington and Manzano}(2003)}]{Carrington03}
\bibinfo{author}{\bibfnamefont{A.}~\bibnamefont{Carrington}} \bibnamefont{and}
  \bibinfo{author}{\bibfnamefont{F.}~\bibnamefont{Manzano}},
  \bibinfo{journal}{Physica C} \textbf{\bibinfo{volume}{385}},
  \bibinfo{pages}{205} (\bibinfo{year}{2003}).

\end{thebibliography}

\end{document}